\documentclass[final,5p,times,twocolumn,authoryear]{elsarticle}

\usepackage{graphicx}
\usepackage{amsmath}
\usepackage{amssymb}
\usepackage{listings}
\usepackage{hyperref}
\usepackage{nameref}
\usepackage{aas_macros}
\usepackage{csquotes}
\usepackage{rotating}
\usepackage{hyperref}
\usepackage{siunitx}
\usepackage{ulem}
\usepackage[usenames, dvipsnames]{color}

\journal{Astronomy and Computing}

\begin{document}

\begin{frontmatter}
\title{Machine Learning on Difference Image Analysis: A comparison of
methods for transient detection}

  \author[iate,oac]{B. S\'anchez}\corref{mycorrespondingauthor}
      \cortext[mycorrespondingauthor]{Corresponding author}
      \ead{bruno@oac.unc.edu.ar}

  \author[iate,oac]{M. J. Dom\'{\i}nguez R.}
  \author[iate,oac]{M. Lares}
  \author[utrgv,utsa, cgwa]{M. Beroiz}
  \author[iate,fceia]{J. B. Cabral}
  \author[iate]{S. Gurovich}
  \author[oac]{C. Qui\~nones}
  \author[oac]{R. Artola}
  \author[iate]{C. Colazo}
  \author[iate]{M. Schneiter}
  \author[iate]{C. Girardini}
  \author[iate]{M. Tornatore}
  \author[dfa,iimct]{J. L. Nilo Castell\'on}
  \author[iate,oac]{D. Garc\'{\i}a Lambas}
  \author[cgwa, utrgv]{M. C. D\'iaz}

\address[iate]{
   Instituto De Astronom\'ia Te\'orica y Experimental (IATE--CONICET),
   Laprida 854, X5000BGR, C\'ordoba, Argentina.}
\address[fceia]{
   Facultad de Ciencias Exactas, Ingenier\'{i}a y Agrimensura, UNR,
   Pellegrini 250 - S2000BTP, Rosario, Argentina.}
\address[cgwa]{
   Center for Gravitational Wave Astronomy, 
   University of Texas Rio Grande Valley, 
   Brownsville, TX, USA.}
\address[utrgv]{
   University of Texas Rio Grande Valley (UTRGV),
   One West University Blvd.
   Brownsville, Texas 78520, USA.}
\address[utsa]{
   University of Texas at San Antonio (UTSA),
   1 UTSA Circle, San Antonio, TX 78249, USA.}
\address[oac]{
   Observatorio Astron\'omico de C\'ordoba, 
   Universidad Nacional de C\'ordoba (OAC--UNC),
   Laprida 854, X5000BGR, C\'ordoba, Argentina.}
\address[dfa]{
   Departamento de F\'{\i}sica y Astronom\'{\i}a, 
   Facultad de Ciencias, 
   Universidad de La Serena. 
   Av. Juan Cisternas 1200, La Serena, Chile.}   
\address[iimct]{
   Instituto de Investigaci\'on Multidisciplinario en Ciencia y Tecnolog\'{\i}a, 
   Universidad de La Serena. 
   Benavente 980, La Serena, Chile.}

\begin{abstract}
We present a comparison of several Difference Image Analysis (DIA) techniques, 
in combination with Machine Learning (ML) algorithms, applied to the 
identification of optical transients associated to gravitational wave events. 
Each technique is assessed based on 
the scoring metrics of Precision, Recall, 
and their harmonic mean $F1$, measured on the DIA results as standalone techniques,
and also in the results after the application of ML algorithms, 
on transient source injections over simulated and real data.
This simulations cover a wide range of instrumental configurations, as well
as a variety of scenarios of observation conditions, by exploring a
multi dimensional set of relevant parameters, allowing us to extract
general conclusions related to the identification of transient 
astrophysical events.

The newest subtraction techniques, and particularly the 
methodology published in \cite{zackay_proper_2016} 
are implemented on an Open Source Python package, named \textit{properimage}, 
suitable for many other astronomical image analyses. 
This together, with the ML libraries we describe, provides an effective 
transient detection software pipeline. 
Here we study the effects of the different ML techniques,
and the relative feature importances for classification 
of transient candidates, and propose an optimal combined strategy.
This constitutes the basic elements of pipelines that could be applied in 
searches of electromagnetic counterparts to GW sources.

\end{abstract}


\end{frontmatter}



\section{Motivations: Synoptic era scenario
}
Synoptic sky surveys are promoters of a new era of observational astronomy, 
where data volume is becoming a major challenge, and discoveries are happening
at a rate never experienced before.
Several collaborations, involved in observational astrophysics projects,
are pushing towards a \textit{data-driven} science paradigm, and transforming astronomy. 
The Large Synoptic Survey Telescope \citep[LSST, ][]{ivezic_lsst_2008, lsst_lsst_2009}
 is going to bring this phenomenon to a higher level, where raw data disk-space
consumption is going to be in the PetaByte-scales by the end of the project.
To face this transformation, astronomers have been involved in information technology development 
for several decades, bringing to existence organizations such as 
IVOA\footnote{http://ivoa.net/}, an international alliance committed to 
organize and make available a living archive of historical astrophysical data.

In this context, a new era of observational astronomy is arising since
the first direct detection of Gravitational Waves \citep[GW, ][]{PhysRevLett.116.061102}. 
This historical discovery places a huge responsibility on synoptic telescopes:
the search for the electromagnetic (EM) counterparts of these GW events.
The Transient Optical Robotic Observatory of the South 
(TOROS\footnote{\url{https://toros.utrgv.edu}}), is a project aimed
at identifying those GW sources.
During \textit{Advanced LIGO science run O1} TOROS participated in the search 
for an optical counterpart to the first GW detection \citep{diaz_toros_2016}, in
an effort to determine the origin of its progenitor.
The theoretical scenario was developed in several articles, such as 
\cite{kasen_opacities_2013, barnes_radioactivity_2016}, where models
predict that a GW event like GW150914 involves a merger between compact 
objects.
This model has three possible cases, featuring binary
combinations of Black Hole and Neutron Stars components.
In case a Neutron Star is one of these, the merger will produce an 
EM emission (or \textit{Kilonova}) that will last a couple of days, 
and will be visible at optical and near-infrared wavelengths.
The search for such an elusive signature is a major challenge, and
in many ways can be described as a race against time.
This objective was reached recently, when the event GW170817 was identified by
several collaborations as the first observed Kilonova, see for example 
\citet{abbott_multi_2017, diaz_observations_2017} and references therein.

One difficulty involved is the need for a comparison method between
images obtained on different epochs, since a fast detection of small 
variations in brightness, over a large region of the sky 
is critical to identify the signature of a Kilonova event. 
Another issue is the requirement of a wide field of view, since the instrument 
dedicated to the search would need to cover several hundred square 
degrees per night, and also reaching deep magnitude limits.
If we combine these two simple conditions
of wide sky coverage and temporal resolution
we have that the image 
comparison method will need 
to deal with va\-ria\-ble Point Spread 
Functions (PSF) across several square degrees and, 
at the same time, be able to detect the 
magnitude variations with high fidelity.
One of the approachs to this task is to compare the 
detected sources, and their brightnesses, on each epoch, 
and to select as possible transient candidates any mis-match.
Though this methodology can find difficulties when applied
on crowded stellar fields, or when the transient event is buried
in a galaxy, and its flux is entagled with the luminosity profile
of the host, (this was particularly the case for the event of GW170817).
In the former case the angular cross match of sources can 
have a computational cost which would introduce an undesired time
overhead for a transient discovery survey with high cadence.
In the latter case in order to measure the flux and position
of the transient source a correct modeling of the 
host galaxy luminosity profile must be applied beforehand.

There are several works that tackle this problem by using 
an image subtraction methodology, such as 
\citet{alard_method_1998,  bramich_new_2008}.
This image differentiation approach avoids the
discussed issues of catalog cross-matching.
A different approach has been taken by \citet{zackay_proper_2016} in 
a series of three papers that derive an improved treatment of astronomical 
images from an statistical point of view. 
This is more general that previous works, and translates several 
astronomical common methods such as source detection into statistical language. 
The authors of this method claim that it also reduces the necessity
of a posterior Machine Learning (ML) analysis.            
The astronomical community has been increasing the implementation of ML, 
seizing its capability for solving data processing issues based on 
handcrafted examples \citep{2012amld.book..213F}, 
specially in the transient detection area 
\citep{2010fym..confE..32D, lau_palomar_2009, rau_exploring_2009}.

In this work we implement three difference image analysis techniques:
Alard \& Lupton's (from now on A), Bramich's (B), and  
Zackay's methodology in two separated ways (Z and S)
to simulated and real data.
Afterwards we train ML algorithms to identify interesting 
targets buried in a sea of \textit{bogus} detections, with as extreme 
ratios as 1\% or less. 
The aim of this article is to develop and establish the best possible 
combination of difference imaging and machine learning techniques based on a 
comprehensive metric.
Novel promising methods such as those based on \textit{Convolutional Neural Networks} 
\citep{sedaghat_and_mahabal_2017} will be compared in the future, when TOROS
collaboration had produced enough data to train such complex models.
In this small survey context classical Machine Learning algorithms should to be
suitable enough.

In the following section we introduce the difference image and Machine Learning 
techniques to be studied, in section \ref{sec:data} we present the simulated and 
real data sets that we will be used in the analysis.
In section \ref{sec:results} we discuss the results of difference image analysis techniques
previous to Machine Learning algorithm implementation.
After that we perform the feature selection, and 
analyze different ML algorithms, estimate their performance 
for the classification of \textit{real}/\textit{bogus} detected on 
difference imaging results. 
Finally in section \ref{sec:conclusions} we present concluding remarks and future
prospects over this work. 
The software developed and datasets here used are open source, and can be found on 
TOROS public repositories.\footnote{\url{https://github.com/toros-astro}}

\section{Methods}
\subsection{Difference Image Analysis}
\label{sec:dia}
Difference Image Analysis (DIA) is a technique which directly compares two 
images of the same position in the sky, taken at different epochs.  
It is usual that one of these is a co-addition of many previously taken images, and has very
high signal to noise ratio, known as \textit{reference frame} ($R$).
The other image would be the recently acquired \textit{new image} ($N$).
Both images are assumed to be astrometrically aligned and registered, and so a
special kind of subtraction is performed to deliver, after object detection, the
transient candidates. 
The detection procedure in the difference images
is in essence a classification problem (source or background), that gives
as a result detections of transient sources (TS), detection of
artifacts (Ar)
and missed transients (MT).

\subsubsection{Linearized kernel models}
Image differencing goes back to \cite{phillips_registering_1995}
where the Eq.~\ref{eqn:alard_model} linking the new and the reference 
images at position $(x,y)$, is proposed by means of a deconvolution \textit{kernel} ($Ker(u,v)$), 
bound directly to the change in shape of the PSF between both images.
A direct solution would be like Eq.~\ref{eqn:alard_deconv} by solving
in Fourier space for the kernel 
(here $\widehat{A}$ denotes the Fourier transform of A). 
\begin{align}\label{eqn:alard_model}
 R(x, y) &\otimes Ker (u, v) = N(x, y)\\
\label{eqn:alard_deconv}
  \widehat{Ker} &= \frac{\widehat{N}}{\widehat{R}}
\end{align}
%
%
This can become numerically unstable in the case that the PSF of
$N$ is narrower than the PSF of the $R$ image, and also in the
presence of high frequency noise on $R$, 
making essential to the succes of this methodology 
the good quality of the Reference images.
The linearized kernel model was extensively developed in 
\cite{alard_method_1998}, where a 
decomposition in base functions $B_i(x,y)$ for the kernel is proposed:
$Ker=\sum_i k_i B_i(x,y)$, where $k_i$ are the coefficients. 
Given the assumption that every pixel in the images are drawn from a 
Gaussian distribution $\mathcal{N}((R\otimes Ker)(x, y), \sigma(x, y))$ 
-where $\mathcal{N}$ denotes a normal distribution function-,
we can estimate a Maximum Likelihood using the 
following cost function $Q$, equivalent to a chi-square test:
%
%
%
%
\begin{equation}\label{eqn:cost}
 Q = \int_{(x,y)} \left[ N(x,y) -  (Ref(x, y) \otimes Ker (u, v) ) \right]^2 /\sigma(x,y)^2
\end{equation}
The first proposed kernel was a sum of Gaussian functions
modulated by low order polynomials ($p_u(x)$ and $p_v(y)$), like 
the Eq.~\ref{eqn:alard_expansion}.
\begin{equation}\label{eqn:alard_expansion}
 Ker(x,y) = \sum\limits_{n} a_n \ \mathcal{N}(\mu=0, \ \sigma_u, \ \sigma_v) \ p_u(x) \ p_v(y) 
\end{equation}
This model is not versatile enough for complex-structured PSFs, 
and the image subtractions performed with this methodology may present artifacts.
\cite{bramich_new_2008} proposed a more flexible model modification, 
as it treats each pixel from the PSF as an independent value.
This is basically to use \textit{delta} type basis functions
(Eq.~\ref{eqn:ker_delta}), where each one is modulated by a coefficient 
which represents the pixel value located in position $(u_i, v_i)$.
\begin{equation}\label{eqn:ker_delta}
 Ker(u,v) = \sum\limits_{i} k_i \delta(u-u_i, v-v_i)
\end{equation}
The determination of $k_i$ coefficients is performed during minimization of
the function $Q$, that is, during the likelihood maximisation.
This techniques have been applied before on various astronomical analyses such 
as variable star search, and exoplanet search, 
for example in \cite{oelkers_difference_2013, oelkers_difference_2015} just to name a few.
In \cite{bramich2016difference} the author explores further options for kernel
modelling, adding several selection criteria to optimize the subtraction, 
although this is not being tested in this work.

\subsubsection{Zackay formal image treatments}
The proposed image model by \citeauthor{zackay_proper_2016}
comes from a different statistical point of view, as they choose to represent
the pixels of the images as distributions, and attempt to perform 
hypothesis testing on their values.

In the case of the reference image $R$ the
model is as below:
\begin{equation}\label{eqn:model}
R(x, y) = F_R T  \otimes P_R + \epsilon_R
\end{equation}
with $T$ being the \textit{true} image, i.e. taken with a perfect 
infinite telescope, and no atmosphere influence; 
$F_R \in I\!R$ is the 
transparency, which encloses the atmosphere and instrumental absorption, 
and would play the role of a flux zero point;
the $P_R$ is the PSF, 
which should be normalized to have unit sum; and $\epsilon_R$ is 
the background noise with variance $V(\epsilon_R)=\sigma_R^2$, 
assumed to be normally distributed.

This model is suitable for statistical hypothesis testing of the existence
of a new source, since it allows the definition of simple null and 
alternative hypotheses for the new image:
\begin{align}\label{eqn:hip_null}
\mathcal{H}_0 &: N = F_N \ T \otimes P_N  + \epsilon_N \\
 \label{eqn:hip_alt}
\mathcal{H}_1(q, \alpha) &: N = F_N \ (T + \alpha \delta_q) \otimes P_N + \epsilon_N.
\end{align}
The lack of evidence for the null hypothesis $\mathcal{H}_0$ favors the existence
of a point source at position $q$ with flux $\alpha$ in the new image, 
which is affected by PSF $P_N$, transparency $F_N$ and noise $\epsilon_N$.
According to the Neyman-Pearson lemma \citep{Neyman289}, the most
powerful (following the definition given in 
\citep{neyman_pearson_1933}) statistic is the likelihood ratio test:
\begin{equation}
\label{eqn:likelihood_ratio}
 \mathcal{L} (\alpha, q) = \frac{\mathcal{P}(N, R \ | \mathcal{H}_0)}
                                {\mathcal{P}(N, R \ | \mathcal{H}_1(\alpha, q))},
\end{equation}
which can be calculated without prior information on $T$. 
It can be proven that maximising equation \ref{eqn:likelihood_ratio} 
is the same as maximising the statistic $S$ of equation \ref{eqn:s_max} 
\begin{equation}\label{eqn:s_max}
 S:= \frac{log(\mathcal{L} (\alpha, q))}{\alpha}
\end{equation}
and after intermediate calculations
available in the appendix A of \citep{zackay_proper_2016} the expression for the Fourier transform
of this statistic is obtained in terms of Fourier transform of known quantities:
\begin{equation}\label{eqn:S_hat}
\widehat{S} = \frac{F_N \ F_R^2\ \overline{\widehat{P_N}}\ |\widehat{P_R}|^2\ \widehat{N} - 
                    F_R \ F_N^2\ \overline{\widehat{P_R}}\ |\widehat{P_N}|^2\ \widehat{R}}
		    {\sigma_R^2 F_N^2 |\widehat{P_N}|^2 + \sigma_N^2 F_R^2 |\widehat{P_R}|^2},
\end{equation} 
By following definitions presented in \citep{2015arXiv151206879Z} and 
\citep{2015arXiv151206872Z} it is possible to prove that $S$ is the 
cross match convolution of the real difference image $D$ and its corresponding PSF $P_D$
of equation \ref{eqn:diff}.

\begin{equation}\label{eqn:diff_s}
 \widehat{S} = F_D\ \widehat{D}\ \overline{\widehat{P_D}}
 \end{equation}
By algebraic manipulations the expression of each one looks like 
\begin{align}\label{eqn:diff}
 \widehat{D} &= \frac{F_R\ \widehat{P_R}\ \widehat{N} - F_N\ \widehat{P_N}\ \widehat{R}}{\sqrt{\sigma_N^2 F_R^2 |\widehat{P_R}|^2 + \sigma_R^2 F_N^2|\widehat{P_N}|^2}}\\
 \widehat{P_D} &= \frac{F_R\ F_N\ \widehat{P_R}\ \widehat{P_N}}{F_D\sqrt{\sigma_N^2 F_R^2 |\widehat{P_R}|^2 + \sigma_R^2 F_N^2|\widehat{P_N}|^2}}
\end{align}
with the flux based zero point relative to the difference
\begin{equation}
F_D = \frac{F_N\ F_R}{\sqrt{\sigma_N^2 F_R^2 + \sigma_R^2 F_N^2}}.
\end{equation}

For source detection the authors claim that the best option is to
determine the locations where $S$ presents peaks outside $5\sigma$
(the robust $\sigma$).
This is the same as a \textit{p-value} test cut. We implemented this as a
separated technique, being the true source detection method presented
in the already cited work.
Among other features, this $S$ image statistic has correlated
background noise, and thus is not suitable for every astronomical
information extraction. 
To recover this \citeauthor{zackay_proper_2016} derives the
formulation of an $S_{corr}$ image in Eq. 98 from Appendix C, which is
not affected by the noise level in the vicinity of bright sources and
other additional noise components.

\begin{figure}
 \centering
 \includegraphics[width=0.5\textwidth]{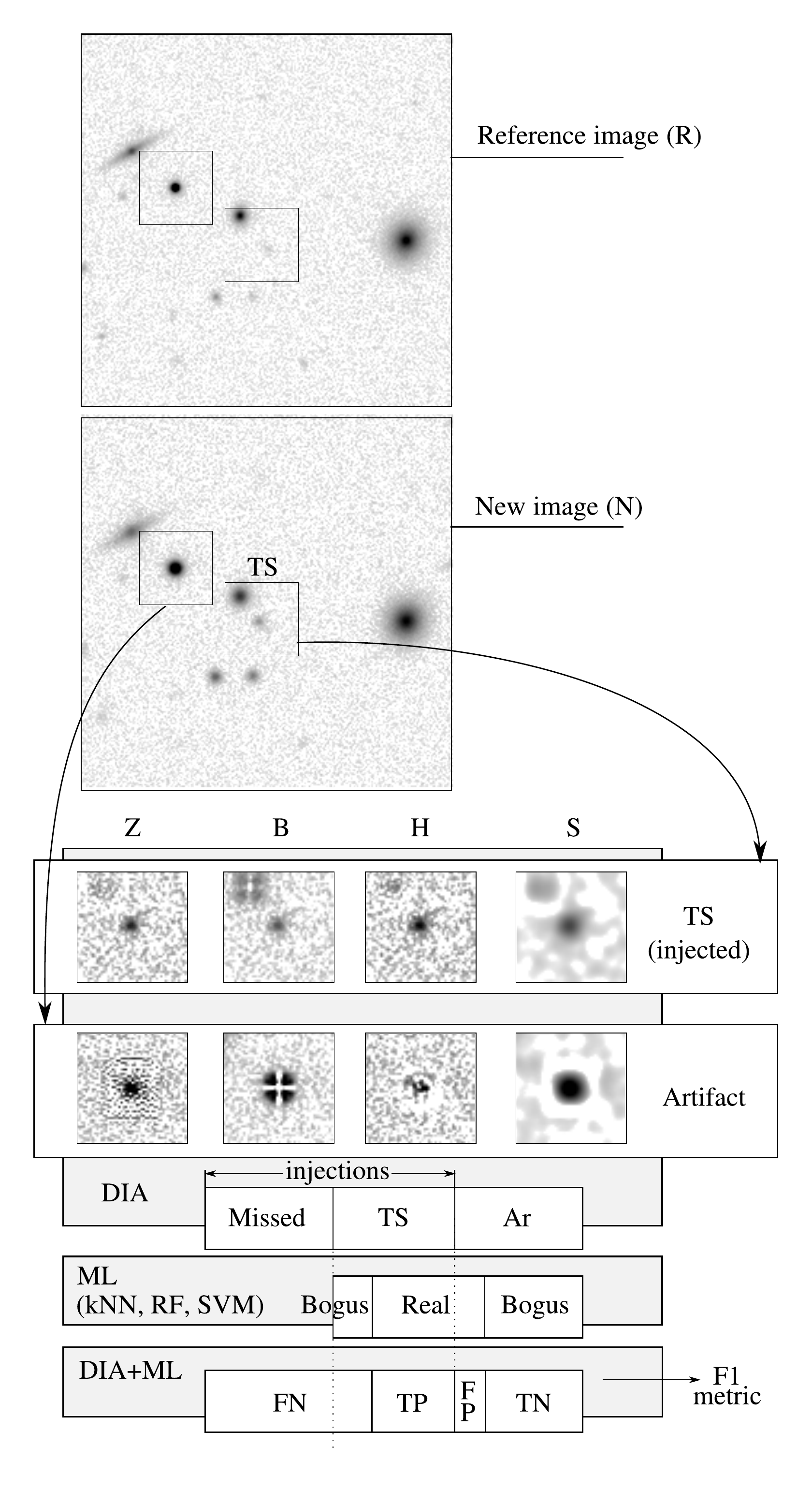}
%
\caption{Diagram showing the combined DIA and ML analysis process
developed in this work.  Reference (R) and new (N) simulated images
are processed using DIA by applying \citeauthor{alard_method_1998} (H),
\citeauthor{zackay_proper_2016} (Z), 
\citeauthor{bramich_new_2008} (B) methods, and $S_{corr}$ (S) image.
We show for comparison, an example of a real object, detected from an
injected transient source (TS) and an example of a bogus object
arising from an artifact (Ar) in the image difference.
In the bottom of this figure, we show three bars, splitted into
several blocks, each one representing different classification results 
at different stages of our multi-stage classification process.
The first bar represents the results of source detection after
the DIA algorithms have performed the image subtraction.
The first two blocks of this bar represent the injected sources, 
which are splitted into Missed and TS detected, and the left block is
the resulting set of artifacts Ar.
The ML algorithms attempt to learn the classification of TS and Ar subsets
into Real and Bogus for each DIA technique.
The results of ML are displayed in the second bar, where the sum of 
the blocks TS and Ar are splitted into Real and Bogus blocks of data.
The combination of the \textit{DIA+ML} results is represented in the 
last horizontal bar, showing the final quantities of False Negatives (FN), 
True Positives (TP), False Positives (FP) and the True Negatives (TN).
As a final figure of merit we calculate the F1 for each combination of
DIA+ML algorithms.
}
 \label{fig:ref_and_new}
\end{figure}

\subsection{Elements of Machine Learning}
\label{sec:mla_intro}
DIA techniques provide the means to detect transient and variable sources on images.
However, as several defects arise and are detected as bogus, it is necessary to classify 
them in order to identify the real sources.
Machine Learning (ML) takes advantage of the power of massive amounts of data 
to generate suitable models to assess the bogus real classification problem.
In the classification of transient objects there are several implementations 
by big collaborations whose capability of collecting data got to overwhelm their
human classification capacity. 
This situation forced them to innovate and apply several machine learning techniques to 
data selection, in order to make manageable their volumes of raw data, and focusing 
their attention on the most promising candidates.
For a summary of the methodologies implemented on recent years
see, e.g., \citet{bloom_automating_2011}.
\subsubsection{Machine Learning Algorithms}
\label{sec:mla_explain}
Machine learning (ML) algorithms rely on the use of data to generalize the relations
between intervening variables in order to make predictions.
There are several classes of algorithms that belong to this area. 
They differ on how they generalize the examples, and more specifically, on 
the way they represent the data as models.
A relevant review on learning with algorithms can be found
in \cite{Domingos:2012:FUT:2347736.2347755}, where a discussion 
on aspects that concern to any ML algorithm is carried out.
The pertinent jargon adopted in this work can be summarized as:
\begin{itemize}
 \item objective class: (also target class) the output variables that we want to predict. 
       In binary classification problems they are commonly referred as ``positive'' and 
       ``negative'' classes.
 \item instance: a data example. Can be a training (``labeled'') instance, 
 		or a new observation.
 \item features: the predictors, in short the measurable and/or computable 
 		quantities that represent  each instance.
 \item score: a value used to quantify the performance of an algorithm to 
 		retrieving the objective class given a training-test dataset.
 \item confusion matrix: a double entry table where it is possible to
 	visualize the classification results in terms of instances correctly
 	and incorrectly classified. 
 	This table originates the score metric values 
        used in classification problems. 
\end{itemize}
For more details or references see, e.g. \citet{Mitchell_1997, hastie01statisticallearning}.
Training datasets are key on supervised ML algorithms, which learn
model representations focused in inferring the objective class 
according to the describing features. A confidence score is
 used to select the best model representation.
In this context, the training data must be labeled, using any previous classification 
available, such as human training.
This validation process is objective at the time of understanding the results, 
and the meta-parameters of the model already trained provide second-order information 
on the data itself.
In this work we employ supervised classification algorithms from the standard
library \textit{Scikit-Learn} \citep{pedregosa2011scikit} --used version is 0.20.1--, 
which is one of the most popular libraries for ML, written 
in the convenient programming language Python.
In what follows we summarize the ML algorithms used in this work.
\begin{itemize}
  \item K-Nearest Neighbors \citep{hastie01statisticallearning}: 
        a really simple algorithm, basically classifies an 
        instance given the most popular kind in its vicinity on feature 
        space, using as a parameter K the number of close instances to
        take into account.  
        Although this method has the advantage of fast training,         
        a drawback is that in a high number of dimensions, euclidean distance
        can be  quite computationally expensive and at the same time inaccurate 
        in terms of the feature space real metric.
        And on top of this you need to store the training data, in order to classify
        new instances.
  \item Support Vector Machines: a quite sophisticated algorithm. In a few words, 
        this technique tries to characterize an hyperplane in the feature space 
        that separates different classes. A major quality of this algorithm is that 
        the procedure to find this hyperplane only depends on inner products of 
        feature vectors (in the linear algebra sense), 
        and so, non-linear transformation kernels can be used to make 
        this classifier able to work in a wider range of problems.
  \item Random Forest (RF) from \citep{breiman2001random}: this is a so called 
        \textit{meta algorithm} since is basically a combination 
        of simpler ML methods. In this case RF is a collection of Decision 
        Trees that use a randomly chosen subset of features to train. 

        A decision tree is a rather simple concept, basically is a chain of 
        \texttt{if--else} that separates the data taking into account only one 
        feature at the time. There exists several variations of this algorithm  
        depending on the statistic that is used to separate the data, such as
        information gain, entropy maximisation, etc. 
        The Random Forest brings an ensemble of trees, that cast a vote, and the majority
        decision is taken. 
\end{itemize}

\subsubsection{Feature selection process}
\label{sec:fsel}
Each feature provides different amounts of information about the
objective class, being more informative features 
the most important ones.
To estimate this relative importance we can use a wide range of techniques that
involve from data visualization, to Principal Component Analysis 
(PCA, \cite{pearson1901lap}).
Since some features might provide redundant information, it is convenient 
to obtain the maximum amount of information with the minimum set of features
since it reduces dimensionality and pruning this non-informative variables 
can lead to an additional computational speed up.
In some cases this can be achieved  by a transformation of the 
feature space, by creating new computable features that reduce 
the dimensionality of the problem (e.g. PCA).
Sometimes this is not necessary since it is possible to discard
redundant features and just use the ones that have better performance.

In order to maximize the performance of the used ML algorithms
we introduce convenient feature selection strategies for 
each case.
%
%
The first strategy is to analyze importance for each feature
individually, this is called univariate analysis.
The simplest technique is just to filter features with low 
variance, since a constant quantity would hold no information regarding
the target class.
This general approach was applied to every tested algorithm.
Another univariate technique is to calculate the mutual information
between each feature and the target class \citep{cover2012elements}. 
Selecting features which maximize this value would in principle 
select the features with higher predictive capability on the target class.
The mutual information technique was used for the KNN algorithm only.

For the decision tree family of algorithms we introduce the 
Random Forest derived feature importance calculation
\citep{Strobl2007, Strobl2008}.
The analysis we used consists of training the model using 
every feature and in the testing stage carrying a random 
permutation of the values 
of each feature, erasing any correlation with the target class.
The decrease in performance of the trained algorithm 
would quantify the importance of the permuted 
feature, without need of re-training the model.
In order to determine the significance of this decrease
we include in the training and testing set a control feature
with random values, 
and compare the importances in relation to it.
Any feature with an importance less than the random feature 
would be discarded.
This technique is biased in the case of correlated variables, 
but this could be avoided by 
pruning them before performing the selection.

Lastly, in the case of SVM we employed a methodology known as Recursive Feature
Elimination \citep{Guyon2002}, which works by using an external 
weighting algorithm, which is evaluated in random subsets of 
features, and recursively pruning those with low weight.
The external weighting algorithm can be any linear model capable of 
delivering a coefficient for each feature, which makes this technique
suitable for linear Support Vector Machines.

\subsection{Evaluation of DIA+ML algorithms performance}
\label{subsubsec:validation}

The focus of this work is centered at the task of recovering
transient sources from telescope images by combining the 
Difference Image Analysis and the Machine Learning methods
in order to maximize recovery completeness and minimize its 
contamination.
To test the training stage of a ML algorithm, a labeled
testing dataset is used to generate predictions, and the performance
can be quantified identically to a hypothesis test, by constructing
the \textit{confusion matrix}, as follows:
(see Fig.~\ref{fig:ref_and_new})


\begin{itemize}
 \item True positives ($TP$) are the injected transient 
    sources correctly detected and classified as \textit{real} instances.
 \item False positives ($FP$) are the artifacts in the image
    differences and the misclassified instances of \textit{bogus}
    objects.
 \item True Negatives ($TN$) are the correctly classified \textit{bogus}
 \item False Negatives ($FN$) are the lost instances due to misclassification
    or missed by the DIA.
\end{itemize}

Notice that the components of the confusion matrix can be computed for
any detection and classification problem, in particular, 
either for the results of the DIA or of the DIA+ML,
changing the previous definitions accordingly.

%
Using the values of the confusion matrix
we can compute more sophisticated scores, useful to
quantify performance metrics for different optimization strategies.
\begin{itemize}
   \item Precision measures how many of the classified as positive
      instances were actually positive.  It can be calculated like
      this: $TP / (TP+FP)$.
   \item Recall ($R$) or True Positive Rate (TPR) characterizes how many of
      the positive examples were actually retrieved.  This is
      $TP/(TP+FN)$.
   \item False Positive Rate (FPR) is the probability of a false detection,
      this is $FP/(FP+TN)$.
   \item False Negative Rate (FNR) is defined as $1-R$, and is the rate
   of lost positive examples.
\end{itemize}

%
%
Precision and recall are useful to check the algorithms performance in
this unbalanced class context, where the bogus or artifact objects rate depends
on the difference imaging method applied.
A more informative value is the $F1$ score, derived from the
precision and recall metrics,
which correctly weights the cost of the
errors of losing transients as well as detecting artifacts.
$F1$-measure is the harmonic 
mean of the $P$ and $R$ metrics,
both equally weighted, and can be used as a final figure of merit. 
It is computed using $2 P R /(P+R)$, which is the same as $2 TP / (2 TP + FN + FP)$
and is also a number from 0 to 1.
This metric is less sensitive to unbalanced classification scenarios, 
because it takes an intermediate value between $P$ and $R$, but
staying closer to the lower value, penalizing the discrepancy of Precision
and Recall.
At the same time is a metric which does not requires the value of the 
$TN$ amount, a quantity we cannot derive from any technique,
due to the nature of the problem.

To asses the performance of a trained algorithm, usually new
data is used. After the training stage, it is possible to detect
cases of \textit{over} or \textit{under} training, by using labeled examples 
which were not processed yet.
For this a standard technique named cross--validation, in which the same
training set is divided into training and testing subsets, is preferred.
This allows to record the mis-classifications and build a confusion
matrix. 
Therefore, we use \textit{stratified} k-fold cross validation
\citep{witten2016data},
which splits the training set in \textit{k} subsets:
\textit{k-1} pieces serving for training and the remaining just for validation purpose.
The results of the validation of this \textit{k} classification algorithms trained 
are the \textit{k} confusion matrices, one for each fold of test data.
The several metrics explained above are then calculated, yielding a
confident performance evaluation of the algorithm.

\section{Simulated and real datasets}
\label{sec:data}
In order to test and compare the different combinations of DIA and ML techniques for transient
detection, we explored a range of different observing conditions using a purpose made data-set
generating simulated images with transients injections. 
However these simulations are not completely realistic, so that we also test the combined 
techniques using observations triggered by GW ALIGO alerts.
In both cases the injection of transients allows to
asses suitable rate estimates to test the performance of the detection methods.
\subsection{Generating a simulated image dataset for ML training}
\label{sec:simdata}
We simulated images using \textsc{Astromatic}
Software\footnote{\url{https://www.astromatic.net/about}},
particularly \textsc{Stuff} and \textsc{SkyMaker} \citep{2009MmSAI..80..422B}, 
which together can produce realistic images of stars and galaxies, for any
given telescope hardware configuration, and including image artifacts
such as saturation, spikes, secondary mirror spider shadows, etc.
Since it is open source software, it is possible to reproduce the results 
of the image simulation by introducing the same configurations.
We simulated the data in several steps:
\begin{enumerate}
\item First we used \textsc{Stuff} to produce a catalog of real objects in the field, 
      including galaxies and stars, containing their positions and real 
      photometric properties, as well as shape parameters.
\item Then this catalog is used to make a fits image using \textsc{SkyMaker}. 
      This is taken as the "reference image" ($R$).
\item Next some stellar sources (transient) are added to the catalog previously created, 
      at random positions, and with random magnitudes drawn from a fixed Luminosity 
      Function (LF) distribution.
\item The final outcome is a "new image" ($N$) with the transients sources included.
\item The last stage of the simulation is to perform the DIA subtraction between $N$ and $R$, 
      and perform the source detection on the resulting difference image.
\end{enumerate}

These steps are repeated once for each point of the explored 
parameter space, having then, one $R$ and one $N$ image 
for each of them.
%
The simulation parameter values cover a relevant range of possible
observational configurations, taking into account three aspects,
namely, the telescope \& site characteristics, the sky stellar
background, and the relative location and brightness of the transient
with respect to their host galaxy. 
The simulations expand eight parameters, and each one has associated
two images, one corresponds to the reference image and the other is
the new image.
The values of the parameters used in the images simulations are
described in Table \ref{tab:parameters}.
Regarding the observational configurations, we considered five
parameters, namely, the diameters of the telescope primary and 
secondary mirrors, the seeing \texttt{FWHM} for $R$ and also $N$, 
the plate scale, and the exposure time.
The values of telescope apertures are selected so that they represent
the available instruments by our collaboration.
The seeing of the $R$ images took values of 0.8, 1, and 1.3
arc-seconds, following empirical determination of TOROS future site
characteristic values (Fig.~ 3 in \cite{renzi2009caracterizacion}).
The seeing of the $N$ images took values of 1., 1.9, 2.5, motivated by
typical and bad observing conditions.
The plate scale and exposure times values are chosen according to the
available CCD cameras.
Regarding the telescope and site characteristics, we include as
particular cases, the Estaci\'on Astrof\'isica de Bosque Alegre
(EABA), the TOROS pilot instrument (TORITOS) and the projected 0.6-m
telescope for the TOROS site.
The number and contrast of the stellar sources are described by the
stellar density parameter.
The range of stellar densities (given by the \texttt{STARCOUNT\_ZP}
parameter) represents fields of different environments going from
typical densities of an mid galactic latitude, and up to densities of
less than 5 degrees from the MW disk center at $l\sim60$ deg of
longitude, with a limiting magnitude of $i\sim19$.
The luminosity distribution of these sources are governed by a power
law, with an exponent which took the values of 0.1, 0.5, 0.9 dexp per
mag (\texttt{STARCOUNT\_SLOPE} parameter in the \textsc{Skymaker}
software)
Also, we allow the variation of the background surface brightness,
using values of 20 and 21 for reference images, typically taken 
mostly on dark nights, and 20, 19, and even 18 for new images, 
taken in different conditions.
%
%
Regarding the host galaxies of the injected transients, we sampled the
relative brightness and the angular distance from the host center from
uniform distribution in the ranges [-4, 1] magnitudes and [0,5]
half light radius, respectively.
This allows to explore different transient/host relative
configurations, including low relative luminosities and position
ranging from the center up to the outer stellar halo.
%
%
For a given observational configuration (or set of parameters), 
we define the magnitude
range where reliable photometry can be obtained, based on the
photometric calibration obtained applying  RANSAC \citep{fischler_ransac_1981} 
robust linear regression on standard sources.
The RANSAC method prunes spurious sources, and obtains an estimation
of slope and zero point values, not sensitive to outliers, and  at the
same time also provides a filter mask identifying this outliers.

The explored parameter space may include configurations which are not 
probable. For instance the combination of a $1.54$ mirror, with $300$
seconds exposure time, in a night with a bright background light
(i.e. moon light), a large plate scale and a broad seeing.
A number of this corner cases are present in the explored
configuration space, and in some of this cases simulated 
images appear completely saturated, and in others the photometric
quality is extremely low.
This corner cases have been discarded, leaving a total of 
26205 groups of $N$, $R$ and DIA differenced images.
The results shown in the next sections include all sensible points in the
explored parameter space, which comprises an heterogeneous combination
of image qualities.
Nevertheless, the independent photometric calibration of each image
informs us the range of validity of flux determination on each configuration
making us able to fairly compare results among the whole simulated dataset.

In section \ref{sec:results}, we discuss the general trends that 
result from our analysis.
Although our parameter space do not cover all possible configurations
for telescope optics and site, we provide the codes that allow to
simulated any other observational configuration.

A total of 3272784 transients were injected, 
placed on top of an extended object 
(as expected in the case of Kilonovae) with random angular 
position and distance relative to the host galaxies 
below 5 half light radius.
The simulated galaxies have different morphological types, 
and also have different redshift values,
random orientation and ellipticity and their luminosities are chosen 
according to a Schechter luminosity function.
The $R$ magnitudes of the transient objects are disposed with 
a random offset from the host galaxy, 
drawn from a uniform distribution, between values -4 and +1 magnitudes.

%
\begin{table*}
\begin{tabular}{llllll}
\cline{4-6}
&&  &\multicolumn{3}{c}{TOROS instruments} \\
\hline
\textsc{parameter}  & \textsc{units} & \textsc{values} & \textsc{eaba} & \textsc{toros}  & \textsc{toritos}\\
\hline
aperture of the telescope &[m]           & [0.4, 0.6, 1.54] & 1.54 &0.6&0.4\\
reference seeing \texttt{FWHM} &[arcsec] & [0.8, 1, 1.3] &1.3&0.8&0.8\\
new image seeing \texttt{FWHM} &[arcsec] & [1.3, 1.9, 2.5] &2.5&1.0&1.0 \\
plate scale &[arcsec/pix]                & [0.3, 0.7, 1.4] &&&\\
exposure times &[sec]        & [60, 120, 300] & & \\
\hline
stellar density &[stars per sq deg]      & [4e3, 8e3, 32e3, 64e3, 128e3, 256e3] &&&\\
stellar luminosity distribution exponent&[dexp per mag] & [0.1, 0.5, 0.9] &&&\\
background brightness ($R$) & [mag per arcsec $^2$] & [20, 21, 22] &&&\\
background brightness ($N$) & [mag per arcsec $^2$] & [18, 19, 20] &&&\\
\hline
relative brightness from host& r-band magnitudes& sampled from Unif(-4,1) &&& \\
angular distance from host & half light radius& sampled from Unif(0, 5) &&&\\
\hline
\end{tabular}
\caption{Parameter space for simulated images to be explored for transient detection.}
\label{tab:parameters}
\end{table*}
%

In the Fig.~\ref{fig:ref_and_new} we present a scheme of the results of the
difference image subtraction and the following ML classification of real and
bogus transient sources.
Besides the reference and new simulated images we show the result 
of the subtractions performed for this pair of images.

The subtractions were carried out by three different implementations 
of the techniques introduced above, namely: Zackay, Alard \& Lupton, 
and Bramich algorithms, and we show stamps of bogus and real objects 
for visual comparison.
We also present the resulting $S_{corr}$ image computed as in 
Eq. 98 from Appendix C of \cite{zackay_proper_2016}.
As can be seen in the stamps, the properties of the subtractions vary 
according to the applied methodology.
It is worth noticing the shape and the appearance of the 
same objects after the subtraction has been performed.
In the case of the transient source injected we see that
every technique presents a point source of almost equal size, 
except for the S image, which shows an enlarged light distribution.
This is due to the nature of the S image, which is a convolution of 
the Z image with its own PSF.
In the second row, we show artifacts originated in the same bright
point source. 
The reason this artifact is consistent in every technique is them 
failing to correctly match the photometric properties of both $R$ and 
$N$ images.
In every case we find different structures and these arise because of 
the intrinsic differences among methods.
The artifact in the Z image has a boxy shape due to the PSF determination
methodology, in the case of the B image the kernel matches the center of 
the stellar sources but fails to adequately account for the flux in the 
wings. 
Similarly in the case of H we find that the source is still visible, though
it lacks a clear structure. 
In the S case we see an excess of intensity, with a smooth profile, 
though surrounded by negative pixels, signs of a flux mismatch in the
Z image.
For the implementation of the \citeauthor{alard_method_1998} method we adopted
the publicly available 
\textsc{HOTPANTS}\footnote{http://www.astro.washington.edu/users/becker/v2.0/hotpants.html}
software by \citet{becker_hotpants_2015} (version 5.1.11).
The \citeauthor{bramich_new_2008} implementation is a Python code by the authors,
available at \url{https://github.com/toros-astro/ois} (version 0.1.14), as well as 
\citeauthor{zackay_proper_2016} implementation, which is also a Python code by the 
authors available at \url{https://github.com/toros-astro/ProperImage}.
Both implementations are built upon standard scientific libraries such as 
NumPy (version used here is 1.15.14),  
SciPy \cite{scipy} (version 1.1.0) 
and Astropy \citep{astropy:2013, astropy:2018} (version 3.0.4),.
and run on versions of Python 2.7 as well as 3.6. %
Also \texttt{ois} and \texttt{Properimage} are fully documented and tested, 
and they are Open Source, free to the community to use. 
Many examples and details of the implementation can be found in the documentation
at \url{http://optimal-image-subtraction.readthedocs.io} and 
\url{http://properimage.readthedocs.io}.
This implementation applies a set of pre-processing stages to the images, 
in order to correctly treat the background, bad pixel masking and interpolation, and 
PSF determination.
It is worth noticing that the Zackay implementation works faster when the 
exposure times of the reference and new images are equal. In this case
there is almost no need for a zero point calibration, 
which saves computational time. 
Alard-Lupton implementation is written in C programming language, also
it employes a simpler Gaussian PSF assumption, making this method 
faster than the others by a factor of almost 4X.

\subsection{Injection of transient objects on observed images}
\label{sec:application}

The simulated images previously used are practical for developing the
DIA techniques and generating a training dataset for the bogus/real
classification problem. Also this approach allows to explore the
dependence of the algorithm performance on different observing
settings and transient properties. Nevertheless, this approach is
limited by the simplifying hypothesis. 
In order to take into account the flaws that arise in the observing
process and subsequent analysis, we present in this subsection the
process of injecting transients sources into real observational
images. We used images obtained by the TOROS collaboration as part of
the follow up of the triggers during Advanced LIGO science run O2.
The images were obtained for the gravitational wave event GW170104,
using the Estaci\'on Astrof\'{\i}sica Bosque Alegre (EABA) 1.54-m Newtonian
telescope.
The instrument was set to white light image acquisition, and the CCD
used was a Apogee Alta U16.
Since the observations were performed hours after the gravitational
wave event trigger was received, we had no previous references of the
selected targets.
The ''reference a posteriori`` methodology was implemented, using
images taken over the following months.
The objects were selected by cross-correlating and filtering the
skymap provided by LSC GCN:20364 and the galaxy catalog from
\citep{white_list_2011}, according to the methodology described in the
first TOROS follow up paper \citep{diaz_toros_2016}.
The set of observed galaxies comprise NGC1341, NGC1567, NGC1808,
ESO0555-022, ESO3564-014, PGC073926, PGC147285, at two epochs
separated by eleven months.
The images were reduced using a standard image processing, and then
co-added for each epoch using the \textit{SWarp}
\citep{bertin_swarp_2010} public software.
The procedure for the subtraction analysis is consistent with 
the one performed on the simulated dataset. 
We perform the subtractions using each image twice, once as a
reference, and once as a new image. 
Given the BH-BH merger nature of the GW event
\citep{ligo_gw170104_2017} it was not expected any EM emission
incoming from this source, and in fact no kilonovae were detected
\cite{sanchez_2018b}.
Therefore we decided to inject transients for the intended analysis. 
In order to simulate a transient object with consistent PSF, we
replicated 15 of the actual stars in each frame, inside the true
dynamic range of the images. 
The total number of realizations was
176 using the observed galaxies, yielding a total of 2640
transient injections.
The subtraction requires the images to be registered, to that end we
developed the python package
\textit{astroalign}\footnote{\url{https://github.com/toros-astro/astroalign}}.
This package was inspired by
\textit{astrometry.net}\footnote{http://astrometry.net} but it does
not rely on a prefixed star catalog, instead it aligns two images
comparing the asterisms drawn by the brightest stars in the field.
%
%
These procedures require a complete pipeline for processing and data
management as the \textsc{CORRAL} framework \citep{cabral_corral_2017}.
This is an open source python package which merges a
database connection interface with a Model-View-Controller paradigm,
making building complex experimental designs 
simpler and straightforward, and letting the
framework figure out the multi-processing
itself.
This allows us to write specific processing steps, like those involved in
DIA and ML combined analysis, according to an intuitive data handling
model.
Therefore, we built a similar processing pipeline, using the same
sequence of steps, for both the real and the simulated datasets.
The completeness and contamination of the transient detection by the
DIA pipelines can be increased and reduced, respectively, by the
application of the ML algorithms previously introduced.
These metrics, among others, were used to rank the transient detection agents,
thus allowing to chose the most suitable approach for applications on
TOROS images.

\section{Comparing transient detection agents}
\label{sec:results}

The DIA methods can deliver either direct subtractions (those which return an
astronomical image without any convolution process), or convolved
subtractions, as the $S$ statistic proposed by
\citeauthor{zackay_proper_2016}. 
For the direct difference image methods, the transient candidate
identifications were performed using
\textsc{SExtractor}\footnote{\url{https://www.astromatic.net/software/sextractor}}
(version 2.19.5), with the same configuration parameters.

\begin{figure*}
 \centering
 \includegraphics[width=0.8\textwidth]{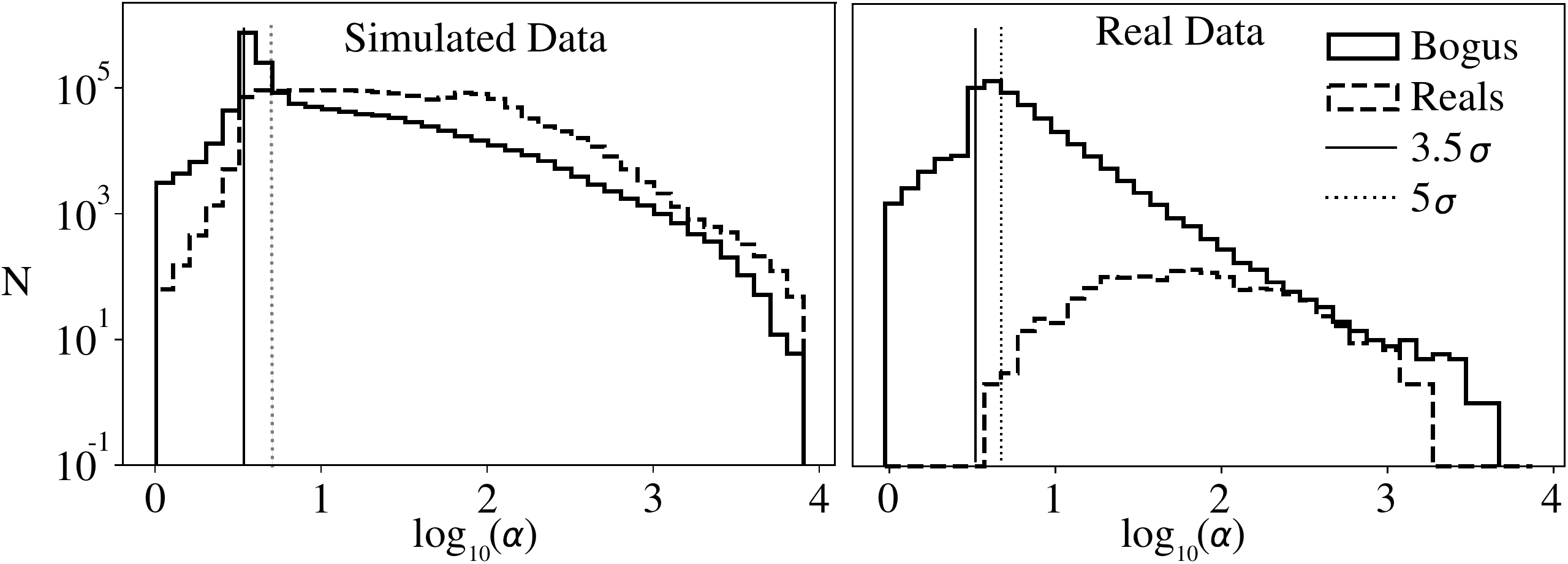}
 \caption{
 Distribution of $\alpha$ values for the artifacts and transient
 sources for $S_{corr}$ detection technique. 
 Notice the logarithmic scales in both axes.
 In vertical lines we include the position of thresholds for $\alpha=3.5\sigma$ and $\alpha=5\sigma$.
 Left: simulated dataset. Right: real dataset.
 }
 \label{fig:sifnificances}
 \end{figure*}

The $S$ image, in turn, is a cross-correlation of the difference image with
its own PSF, and so the detection of transient candidates is different 
from the other DIA methodologies.
By calculating the $S_{corr}$ statistic from the $S$ image and
performing a robust determination of its mean and standard deviation
($\mu$, $\sigma$), we 

can define the significance ($\alpha$) of the detection of a candidate 
in a given pixel as follows:
\begin{displaymath} \alpha = (S_{corr}-\mu)/\sigma 
\end{displaymath}
In Fig.\ref{fig:sifnificances} it can be seen the distribution
of significances for the artifacts and transient sources as
well as both cuts of $\alpha$ for $3.5$ and $5$ in vertical lines, 
for both datasets. 
The chosen threshold of $3.5$ is relaxed with respect to the originally proposed
value of $5$ by \citeauthor{zackay_proper_2016}, attempting to
detect dim transient sources.
Although this increases the number of artifacts, the following ML
analysis is expected to label them as bogus.
Since \textsc{SExtractor} performs a pre--convolution on every
image it scans, we cannot use it on $S_{corr}$ images.
Instead we obtained candidates on the $S_{corr}$ images using 
\textsc{sep}\footnote{\url{https://github.com/kbarbary/sep}}, an open
source package capable of running a source extraction without kernel
pre--convolution.
This software provides photometric measurementes similar to those
delivered by \textsc{SExtractor}.
The threshold used with this source detection technique is again
a limit on $3.5$ over the background. 

Therefore we were able to analyze the candidate sources, its
photometric properties and parameters measured by \textsc{SExtractor}
and \textsc{sep} for every DIA technique, and use this results
as features for the ML stage.
We end having five DIA transient detection results: detections over Zackay 
$D$ images (Z), over Bramich (B) and over Alard-Lupton differences (A), 
the three of them provided by \textsc{SExtractor}; detections over $S_{corr}$
using \textsc{sep} ($S_{\textsc{sep}}$), and detections using a simple 
pixel treshold ($S_{3.5\sigma}$). 
In what follows (Subsec. \ref{SS_diamethods}), we explore the main 
differences between samples of artifacts, real transient sources 
and missed sources for each DIA method.
Since the comparison is based on the very same images, we can 
directly relate the occurrence of both real transient sources 
and missed objects among methods, not being this possible in the 
case of artifacts, since those can be random subtraction
errors misidentified as transient candidates.
The labeled candidates for Z, B, A and $S_{\textsc{sep}}$ DIA results, 
and their photometric quantities measured are the inputs of the feature 
selection process, for training and testing of the ML models aimed 
at doing the bogus--real classification (Subsec. \ref{sec:mla_results}).

\subsection{Performance of the DIA methods}
\label{SS_diamethods}

The fraction of occurrences of the different classes of objects is our 
first piece of information regarding the subtraction methods performance, 
prior to any Machine Learning technique application.
In Table~\ref{tab:fractions} 
we show the number of transient sources (TS),
missed detections (missed) and artifact sources (Ar), along with the
corresponding fractions, TPR, FNR and FPR, respectively.
We also report the F1 statistic after the DIA implementation
that will be later compared to the same statistic after the ML 
application.

By definition the values of FNR and TNR always add up to one, and
FPR can be any number, since the normalization is over the total 
number of injected sources.
Regarding the rates of recovered transient sources (TPR), we read
that there is variability on the results for different techniques.
There is a baseline of 50\% for every technique, and a top value
of 93\%, finding intermediate values in the simulated as well in the
real dataset.
The number of missed objects is larger
in the case of the simulation, since we injected transients in a larger
number of simulated images, covering a wide range of experimental configuration
(as indicated in Table \ref{tab:parameters}).

For the simulated dataset we can read that Bramich finds less transients,
and at the same time produces less artifacts.
For the real dataset it finds more transients than any other technique, 
and produces a relatively low amount of artifacts.
The technique which finds more transient sources is $S_{\textsc{sep}}$, 
with a relatively low number of artifacts for the simulated dataset.
In the case of the real dataset the scenario is the opposite.
The method which generates more artifacts in the simulated case
is Alard-Lupton, yielding more than twice the amount of false 
detections than Zackay technique. 
In case of the real dataset this happens for the $S_{3.5\sigma}$ 
which has a FPR of 200, and Zackay generates less artifacts
than any other DIA method.
This behaviour can be explained with the tendency of $S_{3.5\sigma}$ 
of finding local maximae in the edges of images, or near bright 
sources, an issue we would like to address in the future.

In the simulated dataset we find that the $F1$ statistic is 
systematically higher than the real dataset results.
This is mostly due to extremely high FPR we measure in the latter.
It is clear that we have more sources of confusion in the real images,
this could explained by the presence of instrumental defects on the 
CCD camera used, poor flat field calibration, and correlated noise 
coming from the stacking procedure.
These effects are not straightforward to include in the simulations
therefore we can think of them as a representation of 
an optimistic case scenario, in comparison to the observations.

\setlength{\tabcolsep}{4pt}
\begin{table}
\hspace{-20pt}
\begin{tabular}{lrrrrrrrr}
\multicolumn{8}{c}{Simulated dataset}\\
\hline
{} & TS & Missed & Ar & TPR & FNR & FPR & F1 \\
 \hline 
Z                   &  1,933,065 &  1,339,719 &  3,170,089 &  0.59 &  0.41 &  0.97 & 0.46\\
B                   &  1,596,713 &  1,676,071 &  1,979,948 &  0.49 &  0.51 &  0.60 & 0.47\\
A                   &  1,971,291 &  1,301,493 &  5,537,472 &  0.60 &  0.40 &  1.70 & 0.37\\
$S_{\textsc{sep}}$  &  2,180,390 &  1,092,394 &  2,456,876 &  0.67 &  0.33 &  0.75 & 0.55\\
$S_{3.5\sigma}$     &  2,092,625 &  1,180,159 &  2,700,107 &  0.64 &  0.36 &  0.83 & 0.52\\
\hline
\\
\multicolumn{8}{c}{Real dataset}\\
\hline
{} & TS & Missed & Ar & TPR & FNR & FPR & F1\\
\hline
Z                   &  2296 &        344 &   25914 &  0.87 &  0.13 &    9.8 & 0.15\\
B                   &  2468 &        172 &   47731 &  0.93 &  0.07 &   18.1 & 0.09\\
A                   &  2179 &        461 &  110,025 &  0.83 &  0.17 &   41.7 & 0.04\\
$S_{\textsc{sep}}$  &  2099 &        541 &  128,820 &  0.80 &  0.20 &   48.8 & 0.03\\
$S_{3.5\sigma}$     &  2043 &        597 &  528,927 &  0.77 &  0.23 &  200.4 & 0.008\\
\hline
\end{tabular}
\caption{Number of detections of transient sources (TS), missed
   detections (missed) and artifact sources (Ar), and the
   corresponding rates (TPR, FNR and FPR, respectively) for the implemented 
   DIA methods on the simulated and real datasets, computed before
   ML analysis. 
   We include the $S_{Corr}$ candidates thresholded and 
   extracted with \textsc{sep}.}
\label{tab:fractions}
\end{table}

\subsection{Analyses of the DIA results}
\begin{figure}
\hspace{4pt} 
 \includegraphics[width=0.5\textwidth,keepaspectratio=true]{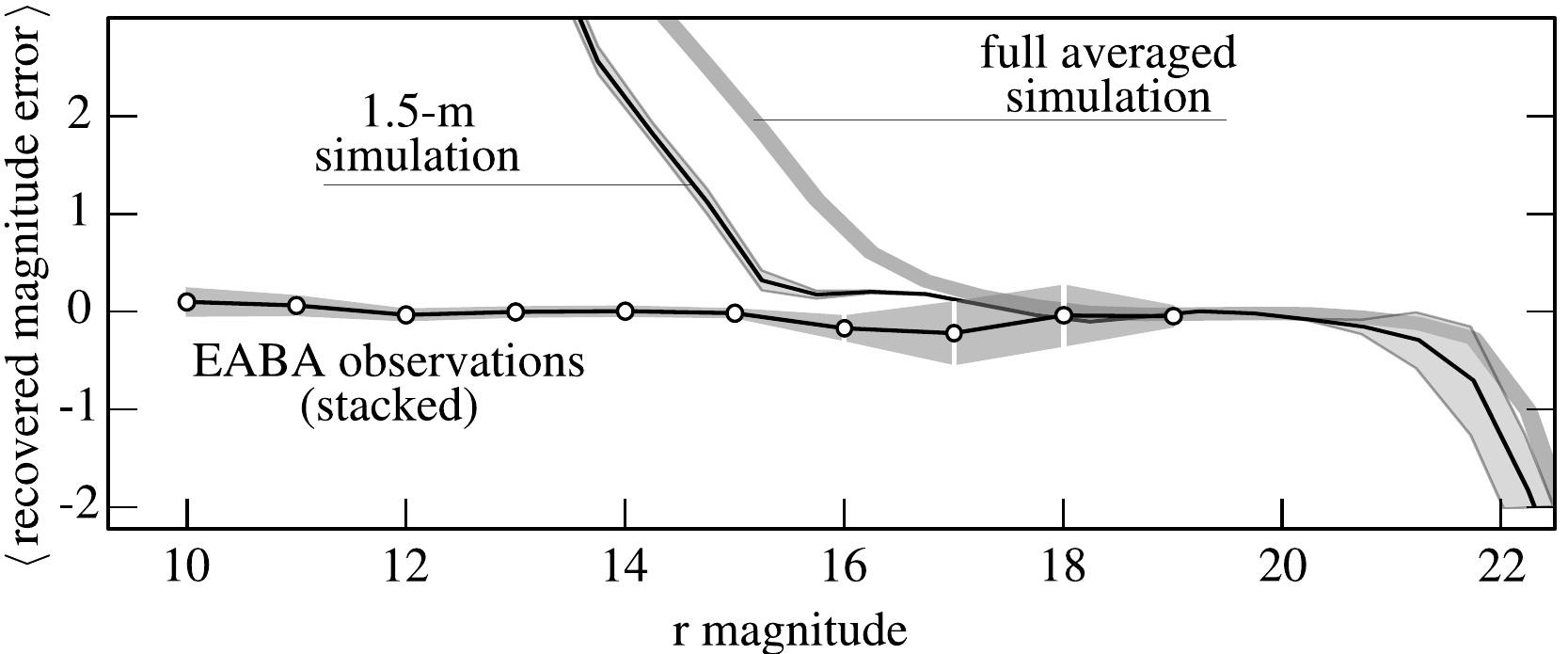}
 \caption{Magnitude difference between injected and recovered sources as a function of
 the magnitude of the injected transients, for the full averaged simulation (thick light gray), the 
 stacked EABA images (circles) and a simulation with an equivalent mirror size and exposure time (solid black).
 Error bars are  $1 \times\sigma$ wide.
 }
 \label{fig:corrected_aper_vs_simulated}
\end{figure}

In order to compare the photometric properties of the transient
candidates for every DIA method, 
we performed a photometric calibration with the flux of the simulated
astronomical sources, by using a robust linear regression as already 
detailed in Sec.~\ref{sec:data}. 
The measured magnitudes of the transients recovered in the simulation
generated images shows 
agreement among the different DIA methods.
We show in the Fig.~\ref{fig:corrected_aper_vs_simulated} the mean and
the standard error of the difference between the injected and measured
magnitudes of the transients recovered in the simulated images, as a
function of the transient r-magnitude.
We also show in this Figure the results for transients injected on
EABA observations, and the subset of simulated images that are closest 
to the observing configuration of the EABA telescope and site.
The difference between the EABA observations and the
corresponding simulation arises because of several factors. 
Most importantly is that the real dataset images are stacks of images, 
and this largely enhances its dynamic range, allowing to observe 
bright sources without saturation, and at the same time, sources at 
magnitudes fainter than 17 with an good signal to noise ratio.
In the simulations, the aperture photometry might not be able to
ideally capture the true flux value of the bright saturated or enlarged
sources.
Prior to reaching the saturation limit, the linearity of the CCD response is lost,
progressively deviating flux measurements from true values.
The limiting magnitude difference between simulations and real data
are due to the actual performance of the EABA instrument, and the 
atmospheric conditions during the nights of O2 follow up observations. 
Taking this into account, it is worth noticing that the simulated and observed images 
are consistent within a range of approximately 5 magnitudes.

\begin{figure*}
 \includegraphics[width=\textwidth]{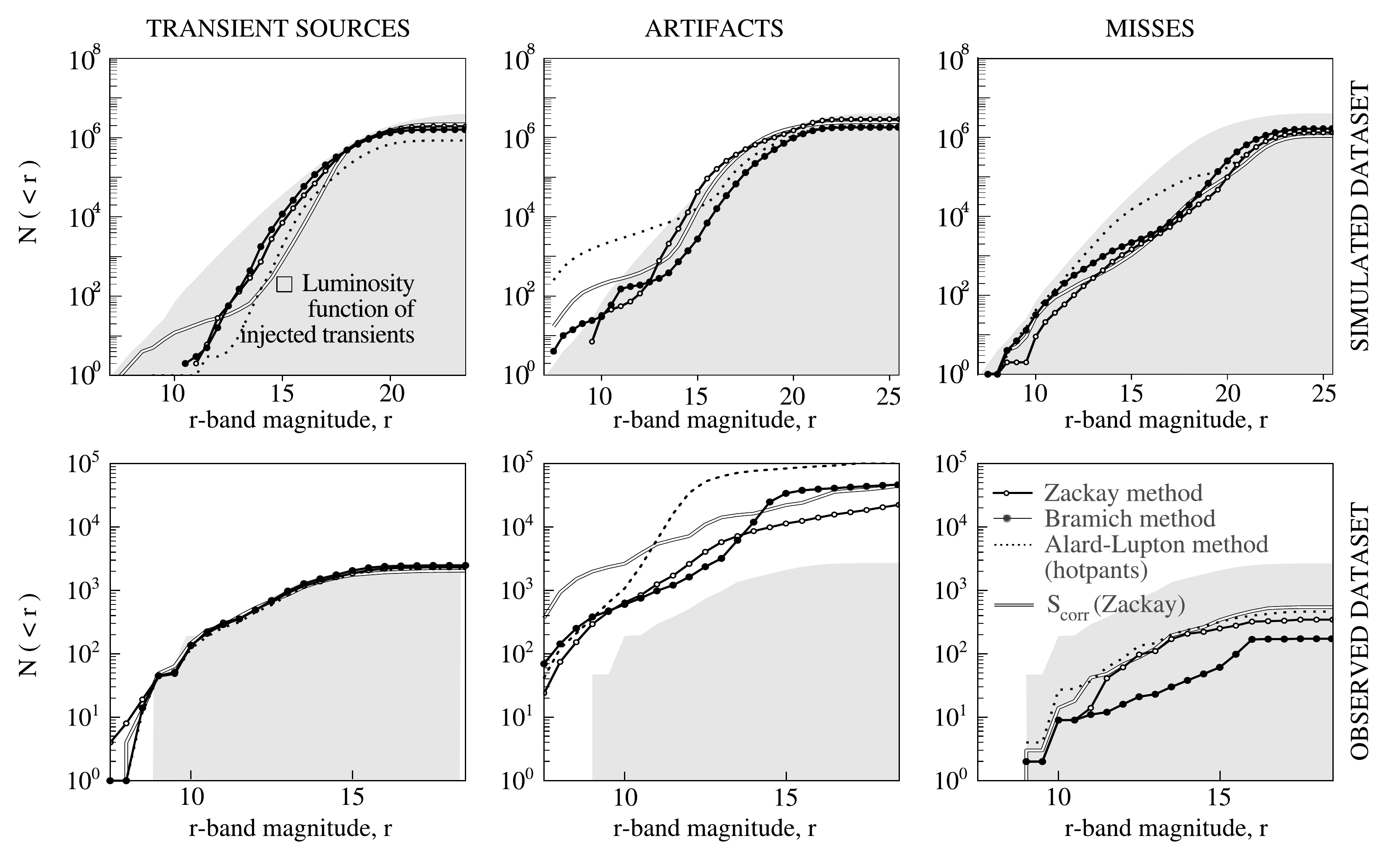}
 \caption{Cumulative Luminosity functions for Transient Sources (TS) 
 (left column), Artifacts (Ar) (center column)
 and Misses (right column) object classes.
 In the first row we have the simulated data and in the second row the real
 dataset. The cumulative luminosity function of the injected sources is 
 displayed in every panel as the shaded area.
 The line code for DIA techniques is: Zackay's is always in lines with white dots, 
 where Bramich's is in lines with filled dots, Alard-Lupton is in dotted lines,
 as well as $S_{corr}$ statistics is in black solid line.}
  \label{fig:lum_functions}
\end{figure*} 

We also studied several statistics for each class of object, trying to gain insight 
into their properties.
The Cumulative Luminosity Function for each class in the simulated dataset
is displayed on the top row of the Fig.~\ref{fig:lum_functions}, 
where it can be seen that for the transient sources the methods are roughly equivalent.
We find that for the magnitude range $r\ge 15$ the methods behave similarly,
but Zackay and S-Corr have more transient sources detected than the rest of the techniques.
In this faint end of the luminosity function the main reason 
for losing objects is the detection limits.
In the bright end $r\le 15$ however, there is not a clear pattern,
and besides the true missed objects, the already discussed errors in the 
magnitude measurements for bright objects due to saturation could
increase the discrepancy with the simulated magnitude values.

In the case of the artifacts, both in the simulated and observed dataset, we find 
similar behaviour in the DIA techniques, although it is worth to 
notice that the accumulated number of objects is in disagree with the 
reported figures in Tab.~\ref{tab:fractions}, this is due to the fact
that many artifact sources present flux measurements without 
astrophysical meaning.
The values of magnitudes calculated for a portion of the artifacts fall 
outside the range presented in the Fig.~\ref{fig:lum_functions}.

The right panel of Fig.~\ref{fig:lum_functions} top row, 
corresponding to the missed objects in the simulations,
shows that all methods fail to recover an important number 
of sources fainter than 21 magnitudes, 
and also objects brighter than 12 magnitudes.
Since this missed, and trasient sources sets of objects are the complement of each other
we find similar explanations for the bi modality of lost sources and the deficits of
injected recovered objects.
Particularly S-Corr is the method which lose less injected sources, 
followed by Alard-Lupton, although the latter losses objects in the whole
range presented, in contrast with the other methods.
As a comparison with the real dataset, we find that there is not such gap in between
the faint and bright end of the missed objects luminosity function, but
instead a smooth ever increasing distribution is observed.
In the real dataset, we have a different picture, with every method
losing objects, almost equally, except for the Bramich method, 
whose distribution seems to be fainter, and shows accumulated numbers
below every other technique.
We also observe a slight increase in the number of objects lost 
for the S-Corr compared to the rest of the techniques, 
although is should be recalled that there is a big gap
between the sample sizes of simulated and real datasets.

In certain way a higher contamination
of bogus may be acceptable since Machine Learning algorithms could 
separate them from the true interesting candidates, and on the other
side, a high FNR is quite undesirable since those candidates are ``lost
forever''.
This compromise should be constrained in advance, and shouldn't be
taken for granted. 
Every algorithm could be optimized by using iterations or
other supplementary techniques, moving the scenario to a more 
favorable ratios scenario.
\subsection{Machine Learning Results}
\label{sec:mla_results}

As explored above, we have photometric properties for the detected candidates
in the several DIA images. On top of this we also have shape properties, as well
as high order statistical moments on their light distributions provided 
by SExtractor -an in $S_{corr}$ case \textsc{sep}-.
This labeled dataset together with the mentioned features, 
constitutes the input instances for training and testing
ML algorithms in the task of classifying the bogus and real
objects.

In order to asses the different scenarios we simulated, our machine
learning experiments were conducted in several steps:
\begin{itemize}
\item We grouped the dataset in terms of three simulation configuration
      values: the mirror diameter of the telescope, the exposure time, and the
      seeing of the new image. This gave us 27 subsets of data, where
      we conducted identical and independent experiments.
\item Each experiment was carried out firstly by splitting the dataset into a 
      training and \textit{final testing} set, with a 20\%-80\% proportion 
      respectively, due to the enormous amount of data available for training.
\item The training subset is used to perform feature preprocess and selection,
      and to perform a k-fold cross validation performance measure for three ML 
      algorithms: k-Nearest Neighbors, Random Forests, and linear Support Vector 
      Machines. 
\item We calculated the confusion matrix for the ML classification (Bogus--Real), and 
      combined it with the DIA performance metrics, in order to rank the
      DIA+ML methodologies using an overall figure of merit.
      This is done by deriving a confusion matrix from the injected sources, 
      through the DIA (Missed, TS, Artifacts), to its final DIA+ML 
      classification results (FN, TP, FP, and TN), such as 
      illustrated in Fig.~\ref{fig:ref_and_new}.
\end{itemize}

The used Machine Learning algorithms as previously stated were:
\begin{itemize}
 \item k-Nearest Neighbors, using 7 neighbors, with uniform weights 
       and euclidean distances (using scaled feature values).
 \item Random Forest, with 800 trees, with up to 7 features per tree, 
       stopping the tree growing if less than 20 examples per leaf, 
       using a Gini impurity criterion.
 \item Support Vector Machines, using L2 norm penalization, with a tolerance parameter
       of $10^-5$, solving the dual optimization problem, and weighting the 
       classes if unbalanced for their frequency.
\end{itemize}
All the configurations for the ML algorithms correspond to Sci-Kit Learn version $0.20.1$.
\subsubsection{The feature selection process}

We performed a preprocessing of the features, by scaling them to 
zero mean, and unit variance.
Afterwards we applied univariated analysis by using
a variance threshold cut of $0.1$, pruning constant features.
Following this simple treatment we used three different feature selection
strategies, adapted to each ML algorithm tested .

To calculate the importances for kNN algorithm, we used the mutual information 
of the features and the target class, selecting the percentile $30$ as a 
threshold cut.

For RandomForest we applied a feature selection process following 
\cite{bloom_automating_2011}, using a RandomForest training stage, and picking 
those features that were the most informative in the majority of the individual trees.
To avoid bias in the selection we
pruned the correlated features, and afterwards we 
followed the methodology described 
in 
associated Python package 
\textit{rfpimp}
in a 10-fold cross validation experiment.
To determine the unimportant features we added to the training dataset a uniformly distributed
random variable.
Using this procedure we can set the zero value of the scale as the importance of this Random
feature. 
We calculated the mean and standard deviation of the 10 values obtained in the 10-fold
experiment, tossing away those features consistent with the values obtained for the Random one.

For Support Vector Machines we applied a Recursive Feature Elimination on a 6-fold experiment, 
using an elimination step of $1$, and choosing $F1$ as the scoring metric to maximize.
%
%

%
\subsubsection{Evaluation of DIA+ML algorithms}

\begin{figure}
\hspace{-20pt}
 \includegraphics[width=0.54\textwidth]{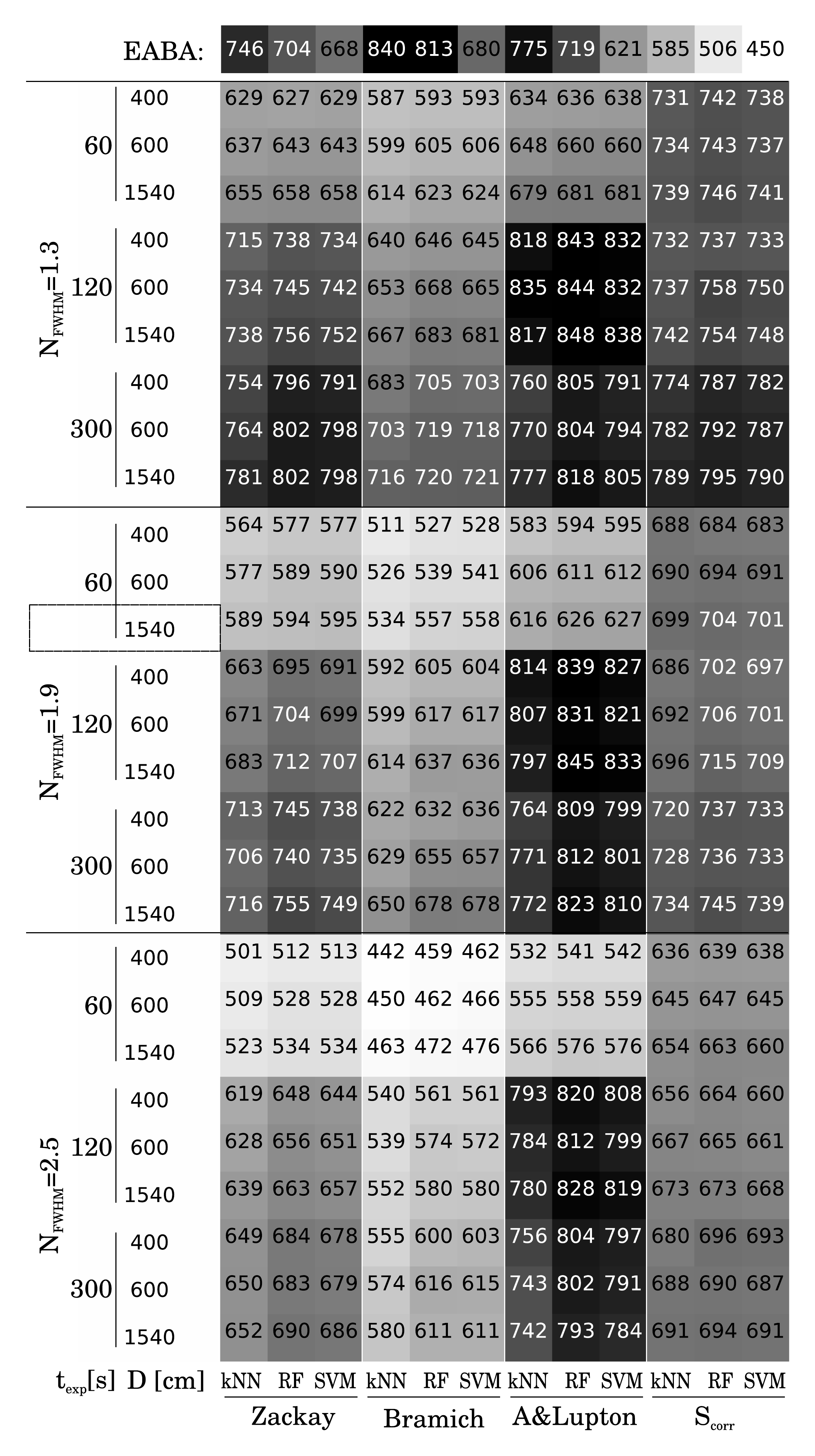}
 \caption{A heatmap of $F1 \times 10^3$ score performance values for each 
 of the 27 instrumental configurations (as rows) where the DIA+ML 
 algorithms have been applied (as columns).
 In the top row the scores for the observed dataset are included in the same grayscale,
 and a separated row is highlighted for comparison, corresponding to the simulations
 with values of ($N_{\texttt{FWHM}}$, $t_{exp}=60s$, $D=1.54$).}
 \label{fig:F1_score_heatmap}
\end{figure}

In Fig.~\ref{fig:F1_score_heatmap} we show a heatmap of values of $F1$ statistic
(scaled by a factor of $10^3$) for the results of the 12 DIA+ML combinations 
(4 DIA techniques and 3 ML algorithms).
The map is obtained by grouping several possible instrumental configurations, 
spanning the dimensions of the \texttt{FWHM} for the new image ($N_{\texttt{FWHM}}$)
measured in arcseconds, 
the exposure time ($t_{exp}$) measured in seconds, and the diameter of the 
primary mirror ($D$) in cm, and performing the ML \textit{train--test} 
experiments on each group.
This covers a wide range of instruments, from small to middle size telescopes, 
and from good to poor observing conditions.
There exists though, several possible configurations which are not covered by our
analysis, designed mostly for TOROS collaboration available instruments.
Nevertheless the analysis is valid for numerous transient search science 
collaborations with instruments falling within the range of our simulation parameters,
such as (piofthesky, black gem, assassn, los alerces, catalina sky survey, etc).
It is also worth noticing that for each of the 27 instrumental configurations
analyzed we are mixing the combination of values of the rest 
of the simulation parameters (listed in Tab.~\ref{tab:parameters}), 
and in consequence including several dissimilar transient detection scenarios
into the same ML experiment.

This result shows an expected dependence in the simulation parameters, clearly 
favouring longer exposure times and smaller seeing \texttt{FWHM} for the new image,
there is also weaker but present dependence on mirror size.
The strongest dependence of the $F1$ value holds with the DIA technique applied.
It is clear that independently of the ML used Alard \& Lupton DIA method has better
performance in terms of $F1$ statistic in the simulations, followed by 
Zackay's techniques (including $S_{corr}$).
There is no major difference among ML algorithms for a given DIA technique, 
but a small advantage seems to be obtained when using RandomForest.

We also include the values of $F1$ for the real dataset in the top of the 
map, which can be compared to the highlighted equivalent simulation.
In the observed dataset we find that generally Bramich is better ranked, and 
the best DIA+ML technique is its combination with kNN.
Notice that the overall $F1$ for this DIA method is much better in this 
observed dataset than in the simulations.
However the range of values for the ML+DIA combination are comparable, 
despite the difference in the $F1$ values of DIA only.

\begin{figure}
\hspace{-20pt}
 \includegraphics[width=0.55\textwidth]{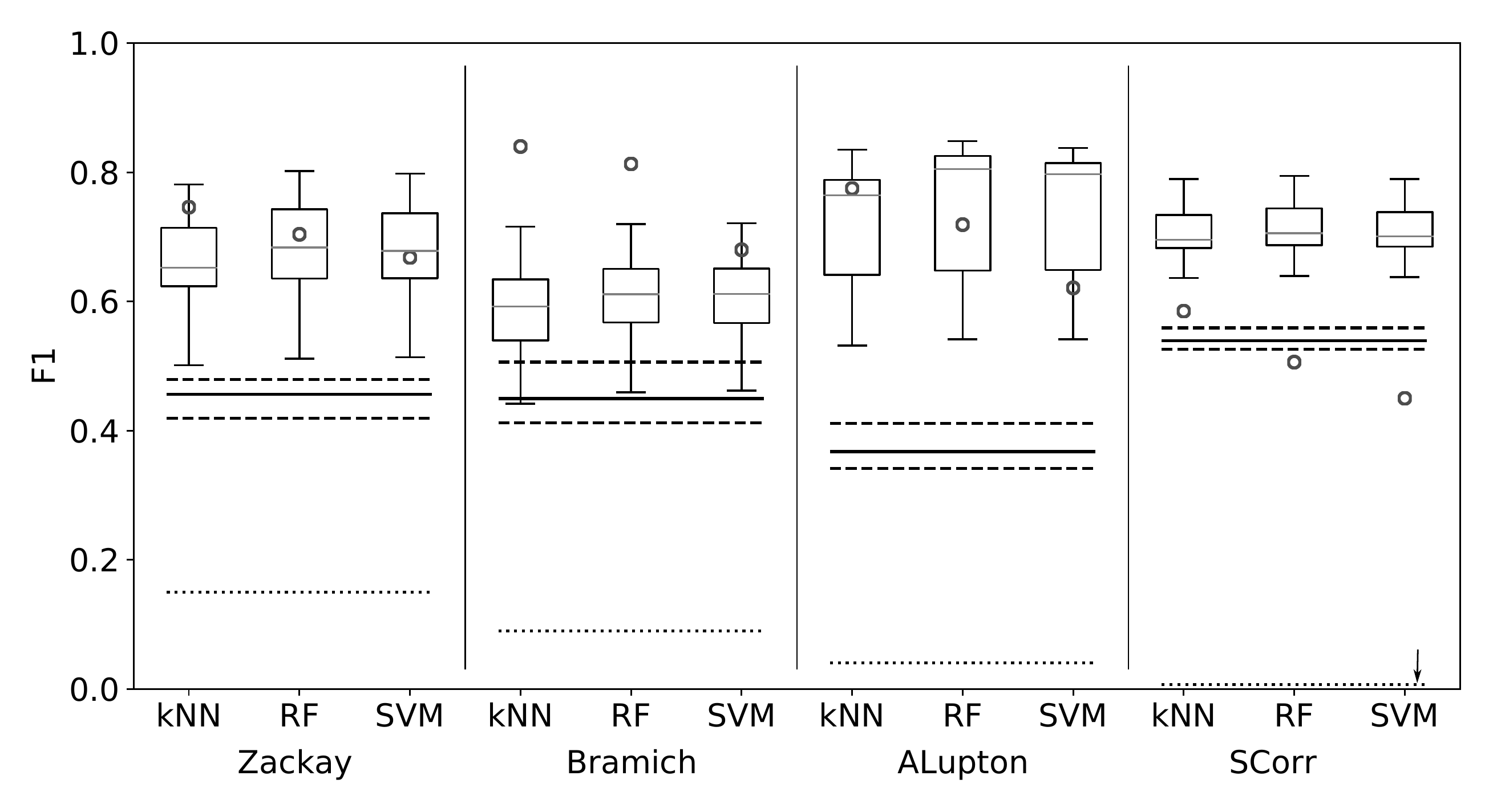}
 \caption{The distribution of F1 values presented in 
 Fig.~\ref{fig:F1_score_heatmap}, displayed as boxes and whiskers.
 Inside the box we have the median, the box edges displays the 25 and 75 percentile, 
 as well as the whiskers show the minimum and maximum values of the sample.
 In horizontal solid lines we show the median of the 
 F1 score prior to ML, and as dashed the quartiles of these values.
 In circles we have the values of F1 for the real dataset. 
 In dotted lines we also show the F1 value for the real data 
 corresponding to Tab.~\ref{tab:fractions}.}
 \label{fig:F1_boxplots}
\end{figure}

In order to better compare the performances reported in Fig.~\ref{fig:F1_score_heatmap}
we have marginalized over the groups of 
instrumental configurations, showing the quartiles of the distribution of $F1$ values 
in the Fig.~\ref{fig:F1_boxplots}.
In order to compare these values before and after ML classification 
we also include as horizontal lines the quartiles of the distribution
of the $F1$ values obtained in each group after the application of 
the DIA methods only.
It is clear from this figure that Alard\&Lupton technique is better ranked
than any other DIA method, and at the same it experiences
the largest boost in performance after the application of any ML algorithm.
We also show in dots the results for the observed dataset and 
in horizontal dotted line we also include the $F1$ value for the 
corresponding DIA methodology (see Tab.~\ref{tab:fractions}).
In the real dataset we find that the combinations which make
use of kNN and RandomForest algorithms are signficantly
better ranked than SVM combinations.
In general Bramich DIA technique results in 
better performance as measured by $F1$ statistic, 
comparable to the best values obtained in the simulations.
However we notice that this result is linked mostly to the 
improvement that ML algorithm provides, which is larger 
than in the simulated dataset.
It is noticeable also, that Alard\&Lupton results
are consistent between simulation and real dataset within the uncertanties, 
being this also consistent with the Bramich results for the real dataset case.
This will be further explored when the TOROS collaboration
instrument data aquisition period begins, and larger
collections of images are available.


\begin{figure*}[h]
\centering
\includegraphics[width=0.9\textwidth]{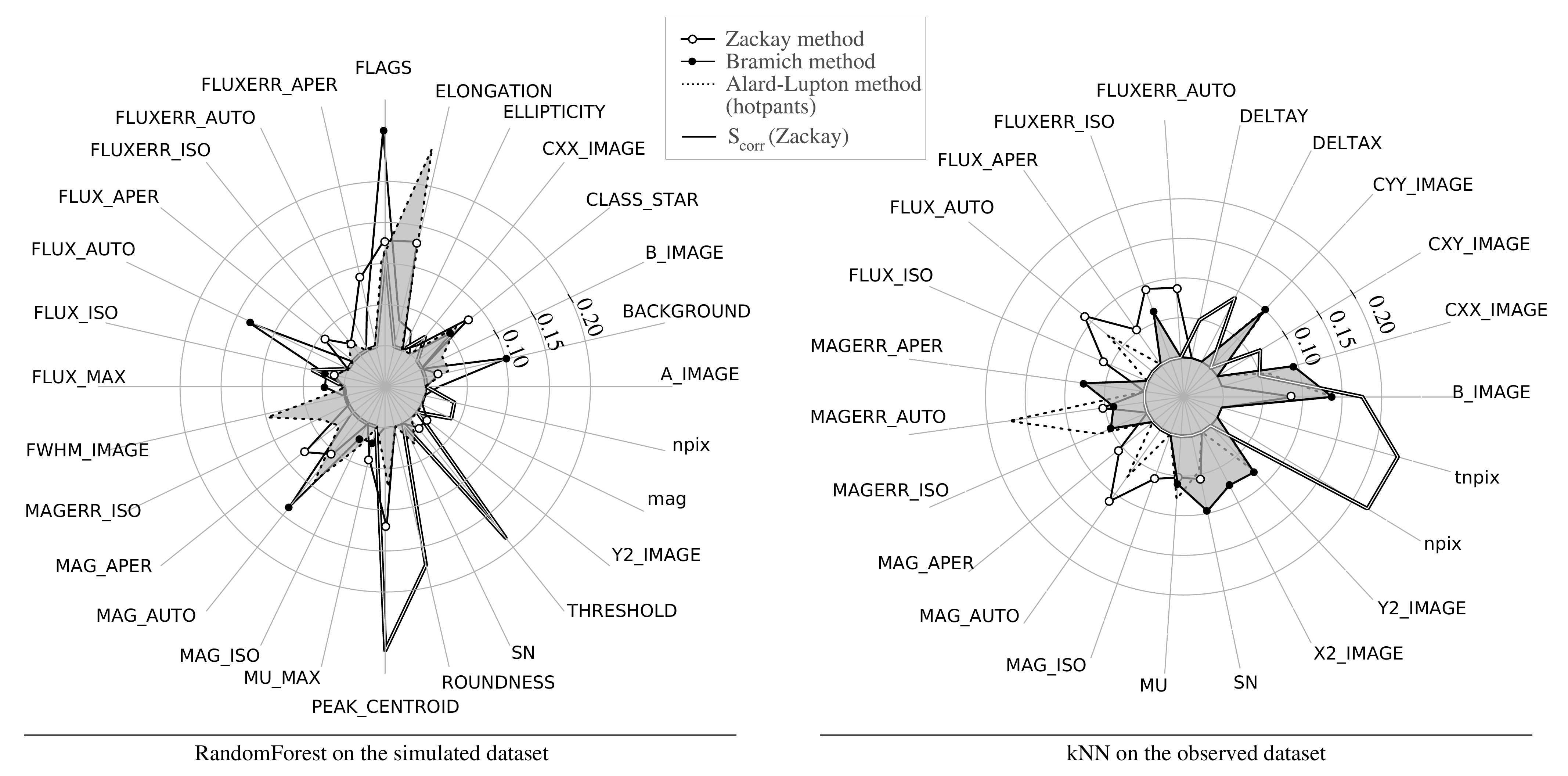}
 \caption{Radial plot of normalized feature importances. 
 The left panel corresponds to the feature selection for the RandomForest 
 algorithm in the simulated dataset. The right panel corresponds to the feature
 selection specific to the kNN algorithm as explained in Sec.~\ref{sec:fsel}, 
 in the real dataset. The details on the calculation and meaning of each feature are included 
in Appendix A. }
 \label{fig:fsel}
\end{figure*}

As a global result we report that the best combination of DIA+ML 
for the simulated dataset is Alard \& Lupton implementation
Hotpants, and RandomForest.
For the real dataset in turn we find that the best performance is
with Bramich DIA combined with kNN machine learning algorithm.
For these two cases we present the selected features and their 
normalized relative importances in Fig.~\ref{fig:fsel}.
The included features in this figure sum up 90\% of the total importance
calculated for the whole set of selected features in each DIA+ML case.
We can see that in the simulation the 5 most important features in the case of
the best ranked DIA+ML for the simulated dataset, that is 
A\&Lupton, are in order \texttt{ELONGATION},
\texttt{FLAGS}, \texttt{FWHM\_IMAGE}, \texttt{MAG\_AUTO} and \texttt{CLASS\_STAR}.
In case of the real dataset, for Bramich+kNN the 5 most important features are
in order \texttt{B\_IMAGE}, \texttt{CXX\_IMAGE}, \texttt{CYY\_IMAGE}, \texttt{SN}, 
\texttt{FLUXERR\_ISO}.
The details on the calculation and meaning of each feature are included 
in Appendix A. 
The disagreement among the maps of importance might be explained considering 
the differences in the DIA+ML algorithm, and its feature selection procedure.
Still, we can detect that the most important features represent as expected, 
shape and brightness parameters.

In order to gain some insights on the simulated dataset, we explored some of the 
dependences of the metrics of performance with the parameters of the injections.
\begin{figure*}[h]
 \centering
 \includegraphics[width=0.8\textwidth]{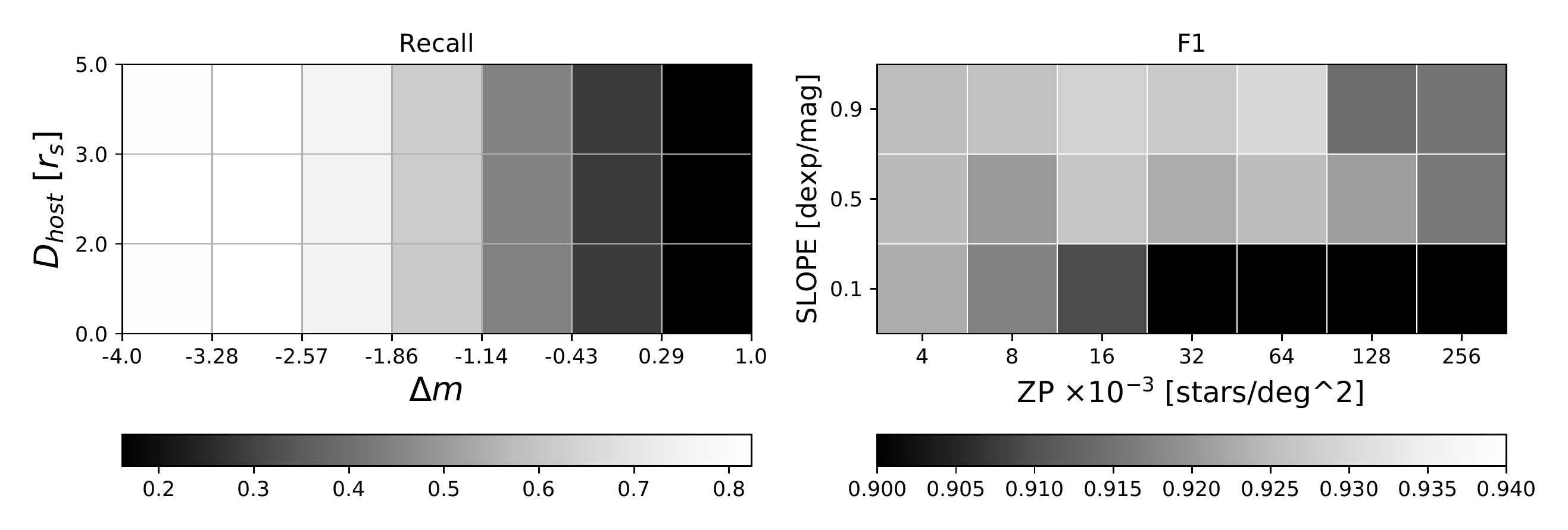}
 \caption{Maps of Recall $R$ and $F1$, for A\&Lupton+RandomForest, for the simulations in two
 different projections. The left panel shows the values of Recall $R$, 
 as a projection of distance of the injected transient
 relative to the center of the host galaxy (in scale radius units) vs the difference
 in brightness between the transient and the host galaxy.
 The right panel in turn, shows values of $F1$ in the plane of slope of the stellar luminosity function
 (\texttt{STARCOUNT\_SLOPE}) on the images, vs the total density of stellar 
 sources (\texttt{STARCOUNT\_ZP}).
 Notice that greyscale is not the same for both panels.}
 \label{fig:maps_alupton}
\end{figure*}
In the Fig.~\ref{fig:maps_alupton} we show the values of Recall $R$ 
and $F1$ metric scores, as two dimensional maps, for two pairs of the simulation parameters, 
for the A\&Lupton+RandomForest technique.
The first one, in left panel, shows the dependance of $R$ 
with the values of relative position and brightness to the host galaxy.
This maps clearly shows a strong dependence of
the Recall of the final ML+DIA on the relative brightness, 
and at the same time shows no dependence on the distance to 
the host galaxy center.
The second map, in right panel, shows the $F1$ values  
as a function of the parameters of the stellar luminosity function of the 
star field on the images: \texttt{STARCOUNT\_SLOPE} and 
\texttt{STARCOUNT\_ZP}.
As explained in Sec.~\ref{sec:simdata} the stellar luminosity function
is a power law, and the exponent is the value of \texttt{STARCOUNT\_SLOPE},
the \texttt{STARCOUNT\_ZP} parameter is the total density of stars in the field
(see Tab.~\ref{tab:parameters}).
This map indicates that in a dense environment the $F1$ is lower, indicating 
the number of stars in the image as source of confusion for the DIA+ML 
technique.
At the same time a lower slope, which traduces into in more bright stars
at a fixed total density, pushes the scoring to lower values.
It is clear then that dense stellar fields, like 
the ones in the galactic plane, are places where a decrese in 
the performance of the DIA+ML is expected.
These class of studies are possible to evaluate by using a multi parameter simulation, 
like the one carried out.

\section{Conclusions}
\label{sec:conclusions}
We developed open source tools for image subtraction following the
DIA techniques from \cite{bramich_new_2008} and \cite{zackay_proper_2016}. 
We also made use of HOTPANTS implementation of Alard-Lupton algorithm.
The Python packages developed, together with HOTPANTS, were mounted on top of
a CORRAL data processing pipeline, in order to systematically perform image subtractions
with transient candidate injections. 
The images used for this were drawn from EM counterpart search, carried out by
TOROS during O2 LIGO-Virgo-Scientific collaboration science observing run in 2017.
This observations were performed during January and November, using the 
``reference a posteriori'' methodology previously applied by TOROS collaboration.
Additionally we developed comprehensive simulations of images, exploring a multi
dimensional parameter space of instrumental configuration as well as 
observing conditions, as detailed in Tab.~\ref{tab:parameters}, 
generating millions of transient injected on top of 
extended sources over several thousands of images.
The nature of the injected transients does not include moving objects,
or stellar variability, focusing the analysis on Kilonova/Supernova
detection scenarios.

The mentioned pipelines measured the ratios of recovery of injected transients, 
as well the source contamination and loss for each DIA algorithm.
We also compared their photometric results, including the $S_{corr}$ image photometry.
In order to separate the true transient candidates from the spurious
artifact detections we applied Machine
Learning algorithms, trained with data generated by our pipeline.
We carried out a feature selection stage, and completed a cross validation train-test experiment,
in order to calculate score metrics such as Precision and Recall, 
useful to compare performances in an unbalanced class context, 
and ranked the methodologies by combining them in the $F1$ statistic.
%

The comparison in the simulated and real data showed that the scenario was of relevance
in the performance of the different combination of methodologies, bringing differences
which can relate to the techniques as well as the nature of data used.
Our results shows that Zackay's image subtraction techniques, 
including $S_{corr}$, are more suitable for transient detection
as a standalone technique. 
However it is clear that the Machine Learning algorithms are key
to complete the task of selecting the relevant transient candidates, 
by setting apart the artifacts which contaminate and are uninteresting.
After the application of these algorithms, and looking at the final performance metric
$F1$ we conclude that, among the DIA+ML combinations tested,
the better ranked technique were: in the real EABA \mbox{images} dataset 
\textit{Bramich+kNN}, and in the simulated dataset 
\textit{A\&Lupton+RandomForest}. 
Although we find consistency, within the uncertanties, among every ML applied technique
in the simulated data, for every DIA method.
This is also valid between the real and simulated dataset, for the A\&Lupton and 
Zackay case.
For these selected methods we also report the more important features, 
determined by Random Forest permutation importance in the case of the simulations,
and by univariated analysis in the case of the real dataset.

Seizing the large and complex simulation generated, 
we analyzed the dependency of the Recall metric on the environment 
of the injected transients, in relation to the host galaxy, 
and the stellar field present in the simulated images.
Concluding that is more likely to generate artifacts and
miss a fraction of transient objects in a dense and 
bright stellar field, and also to miss 
a greater fraction of detections if the contrast in brightness
with the host galaxy is small, independently of its spatial relative location.

These results are very important for the development of the data 
processing pipeline of the TOROS collaboration which comprises 
different telescopes and instruments.
A future extension of this work is to tackle the genuine transient 
time-series astrophysical classification for large amounts of data.

Simulation results and data used in this work 
are available to the community, 
in the format of candidates catalog tables at 
\cite{sanchez_bruno_o_2019_2658714}\footnote{ \url{https://zenodo.org/record/2658714}}.

\section*{Acknowledgements}
This work was partially supported by the Consejo Nacional de Investigaciones 
Cient\'{\i}ficas y T\'ecnicas (CONICET, Argentina) 
and the Secretar\'{\i}a de Ciencia y Tecnolog\'{\i}a de la Universidad Nacional 
de C\'ordoba (SeCyT-UNC, Argentina).
M.B and  M.D. acknowledge NSF support through grant NSF-HRD 1242090.
JLNC is grateful for financial support received from the GRANT PROGRAMS FA9550-15-1-0167 
and FA9550-18-1-0018 of the Southern Office of Aerospace Research and development (SOARD), 
a branch of the Air Force Office of the Scientific Research International Office of 
the United States (AFOSR/IO).
The authors also thank for their kind suggestions and commentaries to D. Bramich and 
B. Zackay.

This research has made use of the
\href{NASA's Astrophysics Data System}{adsabs.harvard.edu/}, 
Cornell University \href{arXiv}{xxx.arxiv.org} repository, the SIMBAD database,
operated at CDS, 
Strasbourg, France.

Also the 
\href{http://www.python.org/}
{Python} programing language and the following scientific libraries: 
\href{http://www.numpy.org}{Numpy}, \href{http://www.scipy/org}{Scipy},
\href{http://www.astropy.org}{Astropy}, 
\href{http://scikit-learn.org/}{scikit-Learn}, and 
\href{https://github.com/parrt/random-forest-importances}{rfpimp}.\\
%
%
\section*{References}
\label{biblio}
\bibliographystyle{model2-names-astronomy}
\bibliography{rbogus}

%
%

\end{document}